\documentstyle[amssymb,aps,prl,twocolumn,epsf]{revtex}

\begin{document}

\noindent
{\bf Reply to Comment on ``Thermal Model for Adaptive Competition in a
Market'':}
In our Letter \cite{mino} we introduced a generalization of the 
minority game (MG) \cite{mg}, called the thermal minority
game (TMG). 
One of the main new features
was allowing for the stochastic decision making
of the agents, controlled by a temperature $T$. 
In the completely deterministic case $T=0$, the original
MG is recovered.

In their Comment \cite{comment} Challet et al.\  claim that: 
(i) the equations of our model reduce to Eqs.(2) of \cite{comment}; 
(ii) Eqs.(2) of \cite{comment} lead to the ``exact solution of the MG''; 
(iii) the crossover to a random value of the variance for $T\gg 1$ 
found in Figs. 2-3 of \cite{mino} is due to short waiting times 
in the simulations. 
Remarks (i) and (ii) are incorrect. Point (iii) is 
true, and highlights even more the crucial role of the temperature in 
the TMG. 
The central claim of \cite{comment}
is that the effects of the temperature in the TMG ``can be eliminated by time 
rescaling'' and consequently 
the behaviour of the TMG is
``{\it independent} of T''. These statements have no general validity.

Challet et al.\  obtained their Eqs.(2) of \cite{comment} 
by averaging  our equations
both over the  noise $\vec{\eta}$ and the individual strategy
distribution. Averaging over $\vec{\eta}$ 
is legitimate, since it preserves
the full macroscopic dynamics for all $d$ and $T$. However,
averaging over the individual strategies as in \cite{comment}
is too naive and misses important correlated fluctuations. 
Besides, this procedure is conceptually wrong: 
replacing $R^*_i$ by its average is equivalent to allowing 
the agents (who have $s$ fixed strategies 
available) to play with {\it any} strategy formed by a
linear combination of the $s$ fixed ones. 
This is not sensible and contrary to the original spirit 
of the model \cite{mg,mino}.

In Fig. 1 we demonstrate the above assertions by comparing the results
of simulations on the TMG with those resulting from the
equations of Challet et al.\  
Not surprisingly, for $T \sim 0$ the approximation leading to 
Eqs.(2) of \cite{comment} works well (at $T=0$ there is no 
stochasticity and the average is equal to the best strategy).
However, as soon as we turn on $T$, this approximation fails 
to reproduce the correct behaviour. 
Clearly, Eqs. (2) of \cite{comment} miss fluctuation effects and 
do not describe the behaviour of our system for all $T \neq 0$. 

In \cite{cmz} Challet et al.\  approximate 
the r.h.s. of Eqs.(2) of \cite{comment} as the gradient of an
effective Hamiltonian $\cal H$ and study the MG by 
minimizing $\cal H$. 
This procedure is predicated on the assumption of equilibration of the
strategy-use probabilities $\pi_i^a(t)$. This assumption is false for
$d<d_c$ \cite{gms}. 
The consequence is that although their method correctly accounts for 
the phase $d>d_c$, it fails completely for $d<d_c$ (see Fig. 1 of 
\cite{cmz}). 
Thus, to claim as Challet et al.\  do in (ii), that they have found  
the ``exact solution of the MG'' is incorrect and misleading. 

Remark (iii) of the Comment is correct: the crossover to a random
variance for large $T$ observed in \cite{mino} is due to finite
simulation times. The time required to reach the steady state for
$T \gg 1$ is of order $NT$. This is a very interesting observation. 
It means that in the phase $d<d_c$, for any {\em finite} temperature 
larger than a critical value $T_c\sim 1$ \cite{mino}, 
the performance of the system will be better than the
original MG, provided that we wait long enough.
In the inset of Fig.1 we show $\sigma$ vs. $T$ for waiting times much
larger than $N T$ for the values of $d$ of Fig.2 of \cite{mino}. 

In other words, what
Challet et al.\  have correctly pointed out in their Comment 
is that {\it any} degree
of stochasticity above a given threshold makes the TMG perform
better than the MG. The claim that ``the collective
behaviour should be {\it independent} of $T$'' is clearly wrong.
The remarkable structure of the
TMG with the temperature begs for further investigation. 

\vspace{0.1cm}

\noindent
Andrea Cavagna$^1$, Juan P. Garrahan$^2$, Irene Giardina$^3$ 
and David Sherrington$^2$

\vspace{0.1cm}

\noindent
$^1$Department of Physics and Astronomy \\
The University of Manchester, M13 9PL, UK

\vspace{0.1cm}

\noindent
$^2$Theoretical Physics, University of Oxford \\
1 Keble Road, Oxford OX1 3NP, UK

\vspace{0.1cm}

\noindent
$^3$Service de Physique Theorique, CEA/Saclay \\ 
F-91191 Gif-sur-Yvette Cedex, France

\vspace{-0.3cm}

\begin{figure}
\begin{center}
\leavevmode
\epsfxsize=2.6in
\epsffile{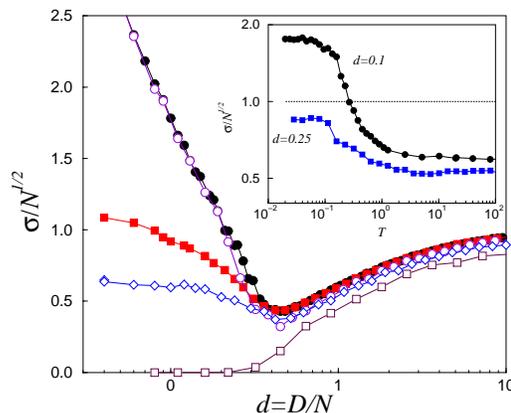}
\caption{
$\sigma$ vs. $d$ in the TMG 
($\bullet \; T=0$,  $\blacksquare \; T=0.32$),
and with the approximation of the
Comment with $\sigma$ calculated as in [4], 
$\sigma^2 = \sum_{ij} \langle \vec{R}^*_i \cdot 
\vec{R}^*_j \rangle$ ($\circ \; T=0.02$,  
$\Diamond \; T=0.32$).
Open squares are
the quantity
$\sum_{ij} \langle \vec{R}^*_i \rangle \cdot 
\langle \vec{R}^*_j \rangle$.
$s=2$, $N=100$, $t=t_0=10^4$, $10^2$ samples.
INSET: $\sigma$ vs. $T$ for $d=0.1$ and $d=0.25$, for waiting time 
$t \sim 10^6 \gg NT$.
}
\end{center}
\end{figure}

\end{document}